\documentclass[a4paper,10pt,journal]{IEEEtran}
\usepackage{cite}
\usepackage{siunitx}
\DeclareSIUnit \GHz {GHz}
\usepackage{graphicx}
\usepackage{amsfonts}
\usepackage{amsmath}
\usepackage[caption=false,font=footnotesize]{subfig}
\usepackage[hyphens]{url}
\usepackage{algorithm}
\usepackage{algpseudocode}

\newcommand{\FF}[1]{{\mathbb{F}}}

\usepackage{tikz}
\usetikzlibrary{arrows,automata}

\usepackage{enumerate}

\newcolumntype{P}[1]{>{\centering\arraybackslash}p{#1}}
\newcolumntype{M}[1]{>{\centering\arraybackslash}m{#1}}
\begin{document}
\title{Secure Data Offloading Strategy for\\ Connected and Autonomous Vehicles}
\author{\IEEEauthorblockN{Andrea Tassi, Ioannis Mavromatis, Robert J. Piechocki, and Andrew Nix}\\
  \IEEEauthorblockA{Department of Electrical and Electronic Engineering, University of Bristol, UK \\ Emails: \{A.Tassi, Ioan.Mavromatis, R.J.Piechocki, Andy.Nix\}@bristol.ac.uk}
}

\maketitle

\begin{abstract}
Connected and Automated Vehicles (CAVs) are expected to constantly interact with a network of processing nodes installed in secure cabinets located at the side of the road -- thus, forming Fog Computing-based infrastructure for Intelligent Transportation Systems (ITSs). Future city-scale ITS services will heavily rely upon the sensor data regularly off-loaded by each CAV on the Fog Computing network. Due to the broadcast nature of the medium, CAVs' communications can be vulnerable to eavesdropping. This paper proposes a novel data offloading approach where the Random Linear Network Coding (RLNC) principle is used to ensure the probability of an eavesdropper to recover relevant portions of sensor data is minimized. Our preliminary results confirm the effectiveness of our approach when operated in a large-scale ITS networks.
\end{abstract}

\begin{IEEEkeywords} Intercept Probability, Secrecy Outage Probability, Data Offloading, Fog Computing, ITS, CAV, V2X.\end{IEEEkeywords}

\vspace{0mm}\section{Introduction and Motivation}
According to the 5G-PPP  and the European C-ITS  initiative, cooperation will be a crucial feature of the future Intelligent Transportation Systems (ITSs), to deliver safety- and mission-critical services among connected vehicles~\cite{anomalyDetectionActions}. 
Future city-scale ITS services will rely upon Connected and Autonomous Vehicles (CAVs) offloading their sensor data onto the Fog Computing layer via a network of Road Side Units (RSUs). The Random Linear Network Coding (RLNC) approach can be used~\cite{8281108}: (i) to improve the reliability of the data offloading process, and (ii) to streamline the removal of duplicated sensor data received by neighboring RSUs. 

CAVs communications are inherently vulnerable to eavesdropping. Traditional physical layer security strategies ensure secrecy by making it impossible for the eavesdropper to recover any of the transmitted packets. 
A CAV, offloading its sensor data employing the RNLC principle, does no longer require a per-packet secrecy. The system can be simplified as~\cite{7214217}:
(i) each transmitted packet is obtained by a linear combination of a number of source packets, and (ii) the source packets can only be recovered after the target number of linearly independent packets has been received.

This paper defines a novel and agile RLNC-based communication strategy. Our approach minimizes the intercept probability of a sensitive data offloading process -- defined as the probability of an eavesdropper recovering relevant portions of the data. With these regards, we will answer the following research questions: \textbf{[Q1]} \emph{What is the impact of RNLC-related parameters on the intercept probability}
and ultimately \textbf{[Q2]} \emph{What is the minimum intercept probability that can be achieved in a large-scale urban testbed?}

\vspace{0mm}\section{System Model and Proposed Solution}
\begin{figure}[t]     
\centering
\includegraphics[width=0.9\columnwidth]{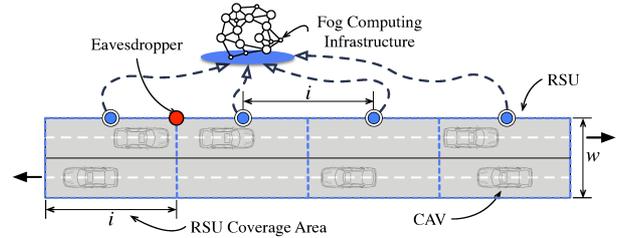}\vspace{-2mm}
    \caption{Considered system model -- A CAV offloads its sensors data onto a Fog Computing Infrastructure according to the proposed RLNC principle.}\vspace{-5mm}
    \label{fig:fogNW}
\end{figure}

We consider the system model shown in Fig.~\ref{fig:fogNW}. $R$ RSUs $\{\mathrm{RSU}_i\}_{i = 1}^R$ are positioned on one side of a straight road section, with width  $w$ \SI{}{\meter}~\cite{6979970}. All RSUs are considered to be ITS-G5 Dedicated Short Range Communication (DSRC) RSUs. We assume that Inter-site Distance (ISD) is the service area of an RSU. Within ISD, an RSU can receive the sensor data offloaded by each CAV with a Packet Error Probability (PEP) smaller than or equal to $\epsilon_\mathrm{R}$. ISD length $i$ is fixed, i.e., each RSU provides coverage across an $(i \times w)$ \SI{}{\meter^2} area. 

We assume that each sensor data stream can be represented as a sequence of packets with the same byte length. Then, we propose to organize the sensor data stream into a sequence of $S_1, S_2, \ldots, S_d$ \emph{source messages}, where $S_t$ (for $1 \leq t \leq d$) consists of $K \geq 2$ consecutive sensor data packets. 
For each source message $S_t$, according to the RLNC principle, a coded packet is obtained as a random linear combination of the sensor data packets forming the same source message -- the random linear combinations are performed over a finite field $\mathbb{F}_q$ with size $q$. A source message can be recovered as soon as $K$ linearly independent coded packets  (associated with the message) are successfully received~\cite{8281108}. RSUs are connected to the same Fog Computing infrastructure that is in charge of: (i) collecting each coded packet, successfully received by each RSU, and (ii) decoding each source message.

\subsection{Secure Data Offloading for Future CAVs}
In our system model, a single eavesdropper is present. In particular, the eavesdropper is stationary and located on the same side of the road where the RSUs are installed. 
Our strategy aims at minimizing the intercept probability by spreading the transmissions of coded packets associated with the same source message across $R \geq 2$ RSUs, related to two or more neighboring coverage areas.
In order to achieve this, we say that each CAV broadcasts one coded packet per source message, starting from $S_1$ and progressively moving to $S_d$. Then the transmission of coded packets restarts from $S_1$. Each CAV broadcasts $N$ coded packets per source message as it drives across $C$ coverage areas -- in the remainder of the paper, $C$ will be regarded as the \emph{reset area}.
In an ITS-G5 DSRC communication systems, Local Dynamic Map (LDM) messages list the locations of all active RSUs in a certain area. 
This list can be broadcast to CAVs by any RSU~\cite{6979970}. Thus, we assume that each CAV is aware of the location, and to that extent of the coverage area, associated with each RSU.

\vspace{0mm}\section{Numerical Results and Discussion}
We considered a scenario where four RSUs provide coverage over a straight stretch of road. In addition, a CAV progressively drives across all the coverage areas offloading its sensor data. As for the channel conditions, we referred to the coverage data collected during the large-scale car trials carried out in the center of Bristol, UK, using our installed experimental ITS-G5 DSRC testbed~\cite{adHocNowCityScale}.

We set the  length $i$  of the ISD to \SI{1200}{\meter} and we place the eavesdropper exactly in between the first and the second RSU (from the left), as per Fig.~\ref{fig:fogNW}. In order to investigate the system performance in the worst-case scenario, we assumed that the channel conditions experienced by the eavesdropper are comparable with those experienced by each RSU.

\begin{figure}[tb]
\centering
\subfloat[$q = 2$]{\label{fig.1a}
    \includegraphics[width=1\columnwidth]{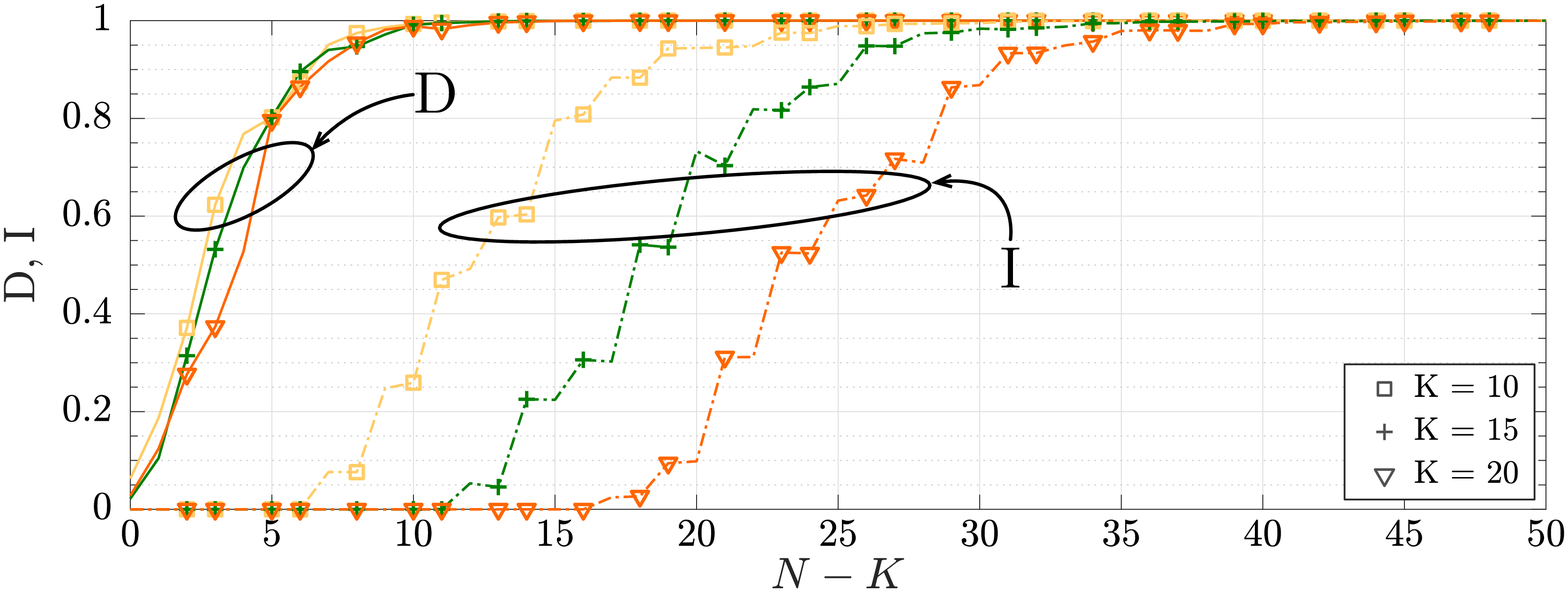}
}
\\
\subfloat[$q = 2^8$]{\label{fig.1b}
    \includegraphics[width=1\columnwidth]{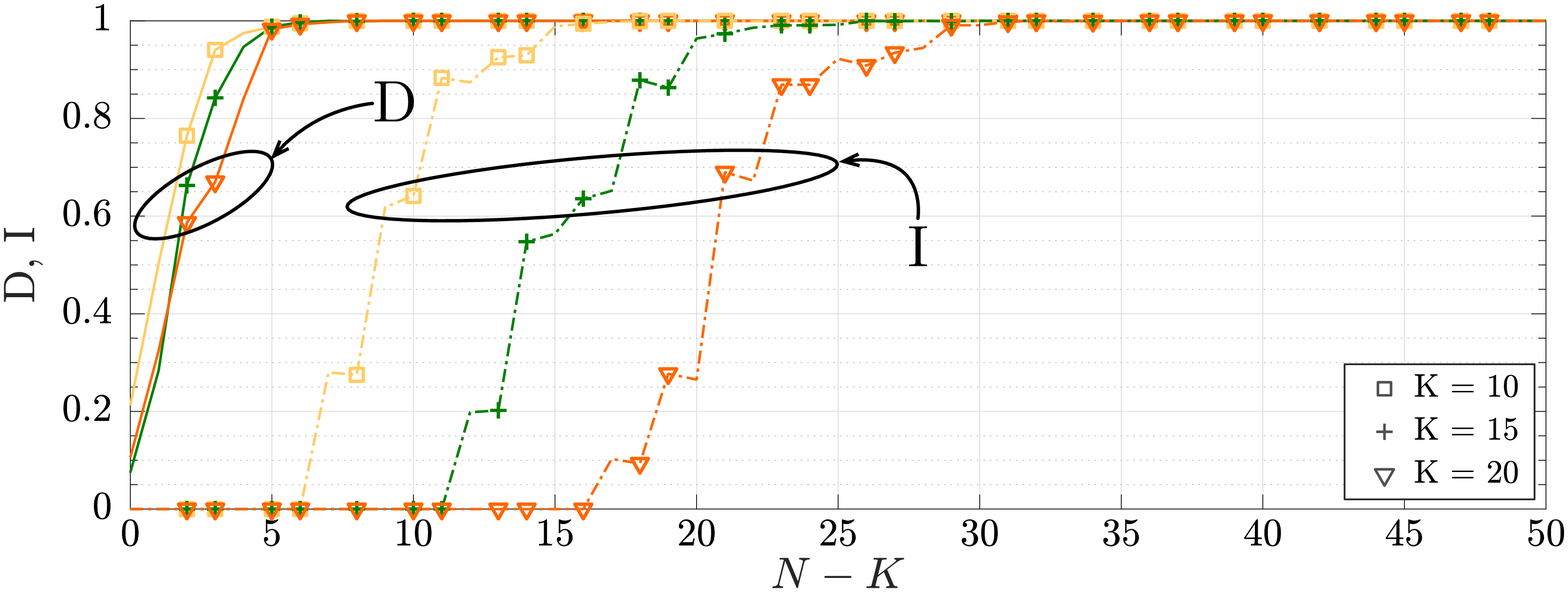}
}
\caption{Recovery and intercept probability as a function of $N-K$, for $q = \{2, 2^8\}$, $K = \{10,15,20\}$, $d = 1$ and $C = 2$.}\vspace{-5mm}
\label{fig.1}
\end{figure}

Fig.~\ref{fig.1} refers to the case where a single source message is considered ($d=1$) and the reset area $C$ is equal to $2$. In particular, this figure shows both the probability of Fog Computing infrastructure recovering a source message $\mathrm{D}$ and the intercept probability $\mathrm{I}$ as a function of $N - K$. Regardless of the value of $K$, both $\mathrm{D}$ and $\mathrm{I}$ increase as the number of the overall coded packet transmissions increases as well, i.e., $N-K$. In addition, due to the increased code efficiency, both $\mathrm{D}$ and $\mathrm{I}$ are sensibly greater as $q$ changes from $2$ to $2^8$~\cite{8281108}.

\begin{figure}[tb]
\centering
\subfloat[$q = 2$]{\label{fig.2a}
    \includegraphics[width=1\columnwidth]{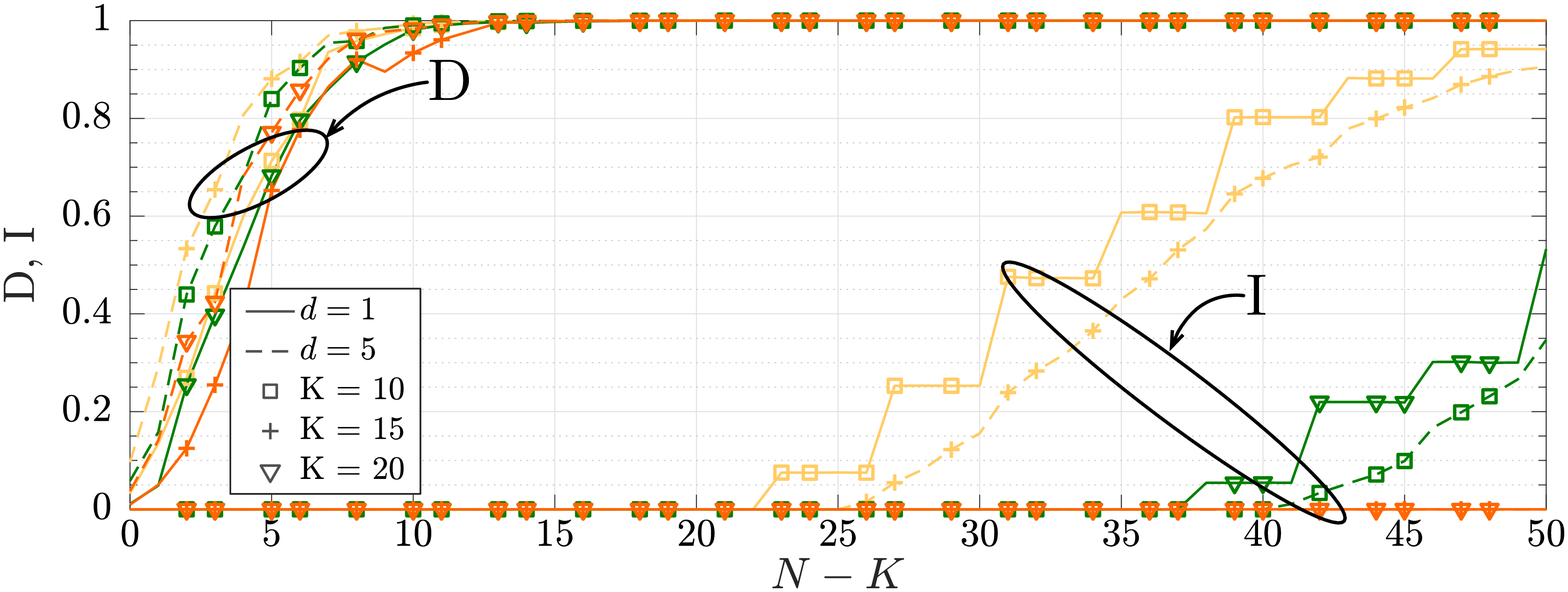}
}
\\
\subfloat[$q = 2^8$]{\label{fig.2b}
    \includegraphics[width=1\columnwidth]{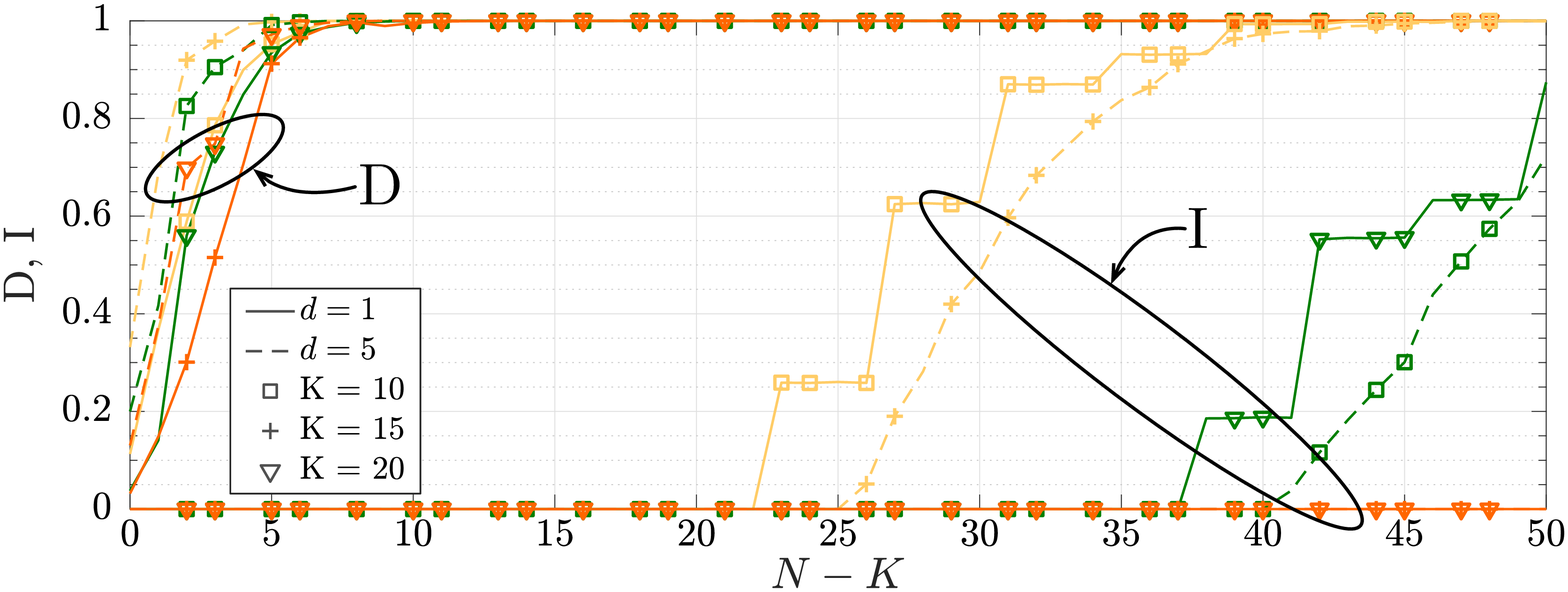}
}
\caption{Recovery and intercept probability as a function of $N-K$, for $q = \{2, 2^8\}$, $K = \{20,30,40\}$, $d = \{1,5\}$ and $C = 4$.}\vspace{-5mm}
\label{fig.2}
\end{figure}

Let Fig.~\ref{fig.1} serve as a benchmark of the overall system performance. In Fig.~\ref{fig.2}, we increased $C$ to $4$. For $d = 1$, we observe that the value of $\mathrm{D}$ remains essentially unaltered -- still ensuring the Fog Computing infrastructure to recover a source message with a probability greater then $0.85$ for $N-K > 5$. On the other hand the value of $I$ is significantly reduced, if compared to the corresponding cases in Fig.~\ref{fig.1}. For instance, for $q = 2$, $K = 10$ and $N-K = 10$, the value of $I$ decreases from $0.64$ to about $0$ as $C$ changes from $2$ to $4$.

From Fig.~\ref{fig.2}, we also observe that the more $K$ and $d$ increase\footnote{For $d > 1$, with a slight abuse of notation, $\mathrm{D}$ and $\mathrm{I}$ signify the probability of the Fog Computing infrastructure recovering a source message and the intercept probability averaged over the source messages $S_1, \ldots, S_d$.}, the more $\mathrm{I}$ decreases, if compared to corresponding cases in Fig.~\ref{fig.1}. This shows how spreading the transmission of each source message across the whole reset area $C$ drastically reduces the intercept probability.

\vspace{-2mm}\section{Conclusions}
We have proposed an agile strategy for securing the offloading of sensor data in a CAV context. As for \textbf{[Q1]} and \textbf{[Q2]}, our numerical results show that the proposed RLNC-based broadcasting strategy ensures the significant reduction of the intercept probability, while the probability of the sensor data being successfully offloaded remains unaltered.

\vspace{-2mm}
\bibliographystyle{IEEEtran}
\bibliography{IEEEabrv,papers}
\end{document}